\documentclass[aps,reprint,superscriptaddress]{revtex4-2}
\usepackage{amsmath}
\usepackage{graphicx}
\usepackage[separate-uncertainty=true]{siunitx}

\newcommand{\avg}[1]{\left< #1 \right>}

\begin{document}

\title{Anomalous tumbling of colloidal ellipsoids in Poiseuille flows}

\author{Lauren E. Altman} \affiliation{Department of Physics and
  Center for Soft Matter Research, New York University, New York, NY
  10003, USA} \affiliation{Department of Physics and Astronomy,
  University of Pennsylvania, Philadelphia, PA 19104, USA}

\author{Andrew D. Hollingsworth} \affiliation{Department of Physics
  and Center for Soft Matter Research, New York University, New York,
  NY 10003, USA}

\author{David G. Grier} \affiliation{Department of Physics and Center
  for Soft Matter Research, New York University, New York, NY 10003,
  USA}

\begin{abstract}
  Shear flows cause aspherical colloidal particles to tumble so that
  their orientations trace out complex trajectories known as Jeffery
  orbits.  The Jeffery orbit of a prolate ellipsoid is predicted to
  align the particle's principal axis preferentially in the plane
  transverse to the axis of shear.  Holographic microscopy
  measurements reveal instead that colloidal ellipsoids' trajectories
  in Poiseuille flows strongly favor an orientation inclined by
  roughly $\pi/8$ relative to this plane.  This anomalous observation
  is consistent with at least two previous reports of colloidal rods
  and dimers of colloidal spheres in Poiseuille flow and therefore
  appears to be a generic, yet unexplained feature of colloidal
  transport at low Reynolds numbers.
\end{abstract}

\maketitle

\section{Introduction}
\label{sec:introduction}

Dispersions of aspherical colloidal particles flow differently than
dispersions of spheres because shear forces cause aspherical particles
to tumble \cite{jeffery1922motion,bretherton1962motion}, and tumbling
influences interparticle interactions \cite{hinch1972effect}.  How
shear-mediated tumbling affects colloidal transport has ramifications
for such diverse application areas as filtration \cite{salerno2006},
drug delivery \cite{blanco2015}, and food processing
\cite{dickinson2015food}.  Tumbling also influences the behavior of
active particles that propel themselves through shear flows, including
motile bacteria and artificial swimmers
\cite{junot2019swimming,ishimoto2023jeffery,omori2022rheotaxis,baker2019fight}.
Despite more than a century of study, the kinematics of colloidal
tumbling are incompletely understood, even for comparatively simple
particle shapes and flow profiles, and even when inertial and
viscoelastic effects may be ignored.

The present study uses holographic video microscopy to explore
anomalous tumbling of axisymmetric ellipsoids in simple shear flows.
Such particles are predicted
\cite{jeffery1922motion,saffman1956motion,bretherton1962motion} to
align preferentially in the plane transverse to the shear direction.
Recent experimental studies, however, show that colloidal rods
\cite{zottl2019dynamics} and bound pairs of colloidal spheres
\cite{altman2021holographic} tend instead to be inclined at
$\theta \approx \pi / 8$ relative to the predicted plane when they are
entrained in plane Poiseuille flows.  Here, we show that prolate
colloidal ellipsoids also tend to be anomalously inclined in steady
Poiseuille flows, in quantitative agreement with previous experimental
studies \cite{zottl2019dynamics,altman2021holographic} and in
qualitative disagreement with theoretical predictions.

\section{Jeffery orbits in Poiseuille flows}

\begin{figure}
  \centering
  \includegraphics[width=0.9\columnwidth]{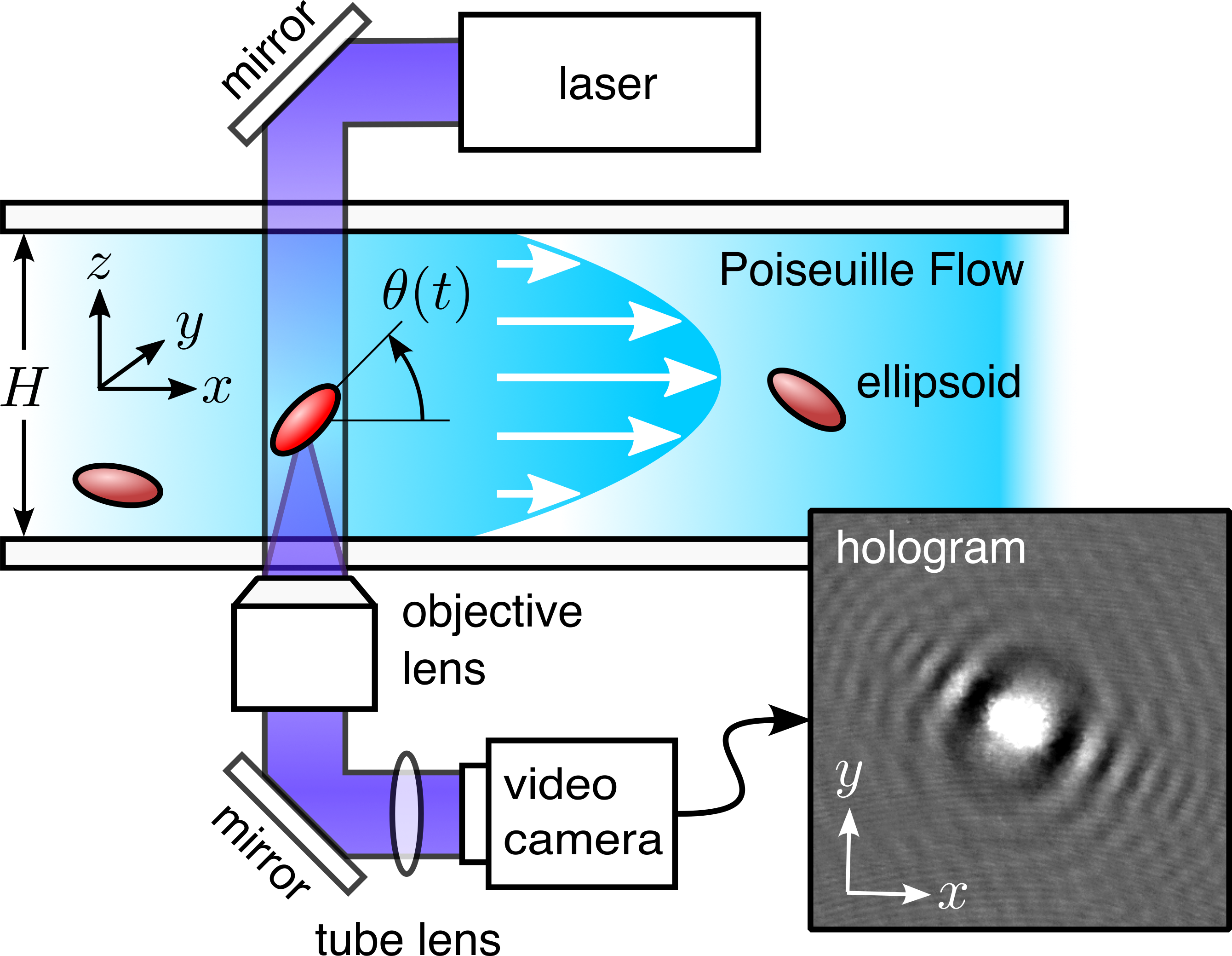}
  \caption{Schematic representation of colloidal ellipsoids tumbling
    as they are transported by the Poiseuille flow in a microfluidic
    channel.  The particles' positions and orientations are recorded
    by an in-line holographic microscope that illuminates them with a
    collimated laser beam.  Light scattered by an ellipsoid interferes
    with the remainder of the beam in the microscope's focal
    plane. The magnified intensity pattern is recorded by a video
    camera. A region of interest from one such video frame captures
    the hologram of a typical ellipsoid, and can be analyzed using the
    Lorenz-Mie theory of light scattering to measure the ellipsoid's
    three-dimensional orientation.}
  \label{fig:schematic}
\end{figure}

Figure~\ref{fig:schematic} schematically depicts the system used for
this study.  An aqueous dispersion of colloidal ellipsoids is
transported down a rectangular channel by a pressure-driven flow.
Particles in the stream pass through a collimated laser beam.  The
light they scatter interferes with the rest of the beam in the focal
plane of a microscope that magnifies the interference pattern and
relays it to a camera.  The image in Fig.~\ref{fig:schematic} is a
region of interest from a typical video frame that captures the
hologram of one ellipsoidal particle. Such holograms can be analyzed
to estimate $\theta$ for each particle passing through the observation
volume \cite{altman2021holographic}.

The Poiseuille flow profile in the channel has a height-dependent
shear rate, $\dot{\gamma}(z) = v_0/z$, that causes the ellipsoids to
tumble as they travel downstream (along $\hat{x}$).  For convenience,
we define $z = 0$ to lie along the midplane of the channel, where the
flow speed is $v_0$.  The flow's vorticity is directed along
$\hat{y}$.  For simplicity, we assume that the height of the channel,
$H$, is large enough compared to particles' dimensions that the shear
may be treated as if it were uniform across the volume of a particle.

The Jeffery orbits of an axisymmetric ellipsoid are most naturally
expressed in terms of the polar orientation, $\theta'$, relative to
axis of vorticity, $\hat{y}$, and the azimuthal angle, $\phi'$, around
that axis, with $\phi' = 0$ being aligned with the gradient direction,
$\hat{z}$.  Neglecting both inertial effects and diffusion, the
orientation of an ellipsoid is predicted
\cite{jeffery1922motion,saffman1956motion} to trace out a trajectory
described by
\begin{subequations}
  \label{eq:jeffery}
  \begin{align}
    \label{eq:jefferyorbit}
    \tan \phi'(t)
    &=
      \lambda \, \tan (\Omega t) \quad \text{and} \\
    \tan \theta'(t)
    &=
      \frac{C \lambda}{(\lambda^2 \cos^2\phi' + \sin^2\phi' )^{1/2}},
  \end{align}
  where $\lambda = a/b$ is the ratio of the major axis, $a$, to the
  minor axis, $b$.  The orientation vector, $(\theta'(t), \phi'(t))$,
  undergoes a periodic orbit at a frequency,
  \begin{equation}
    \label{eq:orbitalfrequency}
    \Omega = \frac{\lambda}{1 + \lambda^2} \, \dot{\gamma},
  \end{equation}
\end{subequations}
that depends on the shear rate and the ellipsoid's aspect ratio.
Different orbits are distinguished by the orbital constant, $C$.
Values around $C = 0$ correspond to log-rolling motion in which the
ellipsoid's major axis is oriented predominantly along $\hat{y}$.
Large values of $C$ correspond to cartwheeling motion in which the
ellipsoid tumbles with its major axis predominantly in the $x$-$z$
plane.

Rotational diffusion causes a Brownian ellipsoid's trajectory to
wander stochastically among orbits with different values of $C$
\cite{hinch_leal_1979}.  For a slender ellipsoid in a simple shear
flow, $C$ is found to be drawn from the probability distribution
\cite{rahnama1993hydrodynamic,rahnama1995effect},
\begin{equation}
  \label{eq:Cdistribution}
  p(C) = \frac{4RC}{(4 R C^2 + 1)^{3/2}},
\end{equation}
which depends on the ratio, $R$, of the ellipsoid's rotational
diffusion coefficients, and therefore on $\lambda$.  Weak inertial
effects at non-vanishing rotational Reynolds numbers destabilize
log-rolling and stabilize cartwheeling, favoring larger values of $C$
than is predicted by Eq.~\eqref{eq:Cdistribution}
\cite{einarsson2015rotation,einarsson2015effect,palanisamy2019}.

Experimental studies of shear-induced tumbling of axisymmetric
colloids
\cite{zottl2019dynamics,einarsson2015tumbling,altman2021holographic}
have been limited by the difficulty of measuring small particles'
translational and rotational trajectories.  The experimentally
accessible angle of inclination, $\theta(t)$, is related to the
orientation angles, $\theta'$ and $\phi'$, by
\begin{equation}
  \label{eq:thetadefinition}
  \sin \theta = \sin \theta' \, \cos \phi'.
\end{equation}
The distribution of inclination angles, $P(\theta \vert C)$, predicted
by Eqs.~\eqref{eq:jeffery} through \eqref{eq:thetadefinition} is
peaked at $\theta = 0$ for all but the smallest values of $C$.  We
therefore expect that the thermal average inclination distribution
\begin{equation}
  \label{eq:jefferytheta}
  P(\theta)
  =
  \int_0^\infty P(\theta \vert C) \, p(C) \, dC,
\end{equation}
will be peaked at $\theta = 0$, regardless of the form of $p(C)$.
Surprisingly, this does not appear to consistent with experimental
observations \cite{zottl2019dynamics,altman2021holographic}, including
those reported here.

\section{Holographic tracking of colloidal ellipsoids}

\begin{figure}
  \centering \includegraphics[width=0.8\columnwidth]{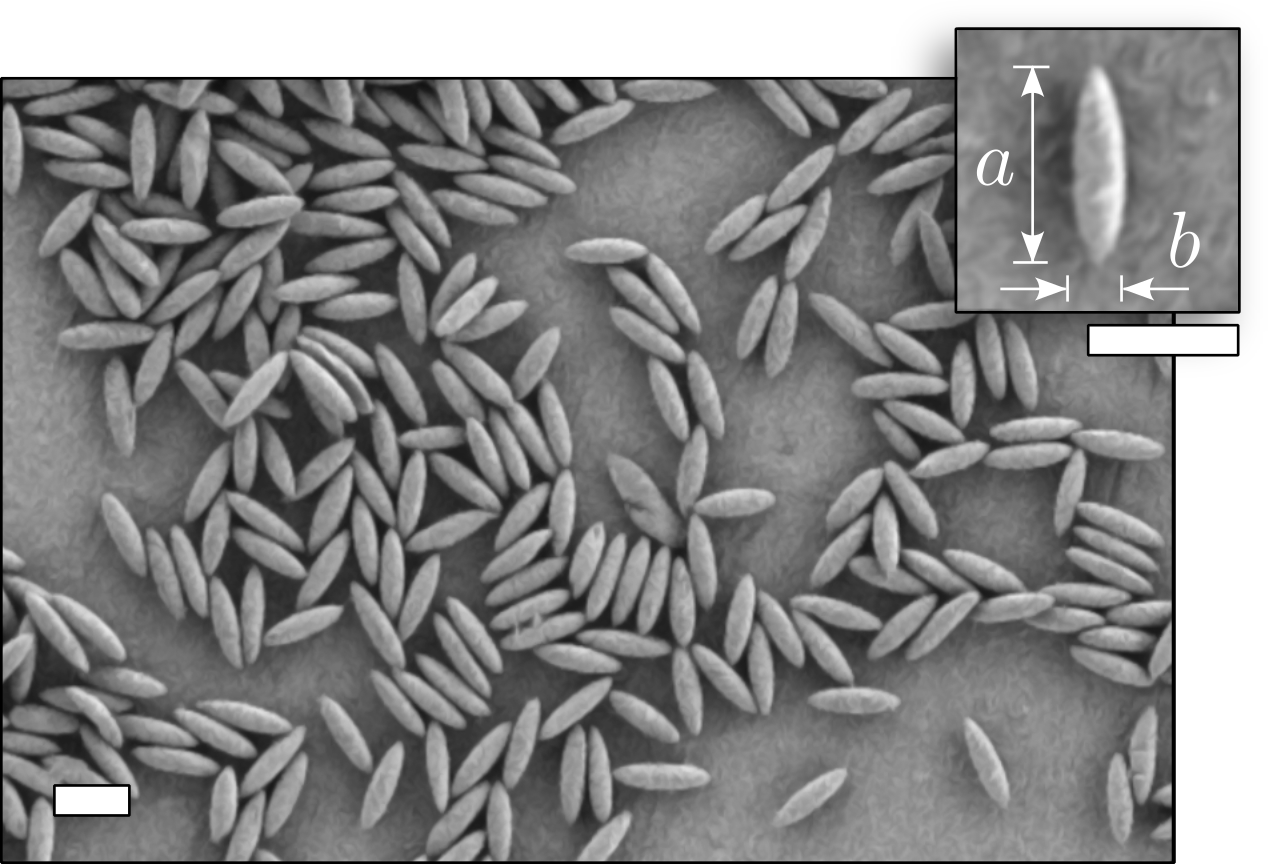}
  \caption{Scanning electron microscope image of colloidal ellipsoids
    deposited onto a graphite substrate and dried.  Inset: typical
    ellipsoid, illustrating ground-truth measurement of the major and
    minor axes, $a$ and $b$, respectively. Scale bars indicate
    \SI{5}{\um}.}
  \label{fig:sem}
\end{figure}

The monodisperse colloidal ellipsoids used for this study are created
by uniformly stretching \cite{mohraz2005direct,klein2013photo}
custom-synthesized polymethyl methacrylate spheres (NYU Colloid
Synthesis Facility, batch CSF02-139-C)
\cite{antl1986preparation,elsesser2010revisiting}.  Scanning electron
microscopy images such as the example in Fig.~\ref{fig:sem} yield a
population-averaged major axis of $a = \SI{4.80(21)}{\um}$ and an
aspect ratio $\lambda = \num{3.96(25)}$.  As has been reported
previously \cite{ferrar2018two}, stretched colloidal spheres differ
slightly in shape from ideal ellipsoids.  They are closer to ideal,
however, than the right-circular rods \cite{zottl2019dynamics} and
bound pairs of spheres \cite{altman2021holographic} that have been
studied previously.

Sterically-stabilized colloidal ellipsoids are dispersed in dodecane
($n_m = \num{1.42}$) at a concentration of
\SI{e6}{particles\per\milli\liter}.  A \SI{30}{\micro\liter} aliquot
is transfered to the input reservoir of a commercial microfluidic
channel (xCell8, Spheryx, Inc.) with a rectangular cross-section that
nominally is $H = \SI{60}{\um}$ high and and \SI{500}{\um} wide. The
10:1 aspect ratio allows us to neglect transverse shear.  The
microfluidic channel is installed in a commercial holographic particle
characterization instrument (xSight, Spheryx, Inc.) that creates a
pressure-driven flow in the channel with a mid-plane speed of
$v_0 = \SI{3(1)}{\mm\per\second}$.  The Poiseuille flow profile has a
shear rate that varies linearly from
$\dot{\gamma} = 2 v_0 / H = \SI{100}{\per\second}$ at the walls to
$\dot{\gamma} = 0$ at the midplane.

\begin{figure*}
  \centering
  \includegraphics[width=0.9\textwidth]{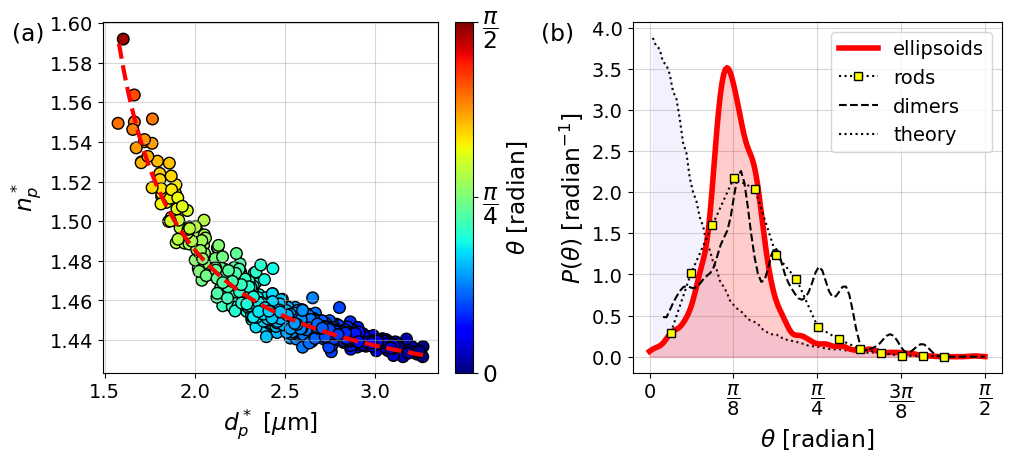}
  \caption{(a) Effective-sphere properties of \num{1712} colloidal
    ellipsoids, with each point representing the holographically
    measured diameter, $d_p^\ast$, and refractive index, $n_p^\ast$,
    of a single particle.  The dashed (red) curve is a fit to
    Eq.~\eqref{eq:parametric}.  Each data points is colored by the
    angle of inclination, $\theta$, associated with its position along
    the parametric curve.  (b) Distribution of ellipsoid inclination
    angles, $P(\theta)$ obtained from the data in (a), compared with
    independent results from for colloidal rods
    \cite{zottl2019dynamics} and dimers \cite{altman2021holographic}.
    The theoretical prediction for Brownian ellipsoids is obtained
    from Eqs.~\eqref{eq:jeffery} through \eqref{eq:jefferytheta}.}
  \label{fig:angledistribution}
\end{figure*}

The instrument analyzes each single-particle hologram \cite{lee07a}
with the Lorenz-Mie theory of light scattering
\cite{bohren83,mishchenko_scattering_2002,gouesbet_generalized_2011}
to obtain the diameter, $d_p^\ast$, and refractive index, $n_p^\ast$,
that describes an effective sphere encompassing the particle
\cite{odete2020role,altman2021holographic,altman_interpreting_2020,abdulali2022multi}.
These effective-sphere parameters are related to an ellipsoid's angle
of inclination through Maxwell Garnett effective-medium theory
\cite{markel_introduction_2016}.  An ellipsoid lying in the focal
plane, $\theta = 0$, has an effective diameter somewhat smaller than
its major axis.  Because the actual ellipse fills only a fraction of
this enclosing sphere, however, its effective refractive index in this
orientation is only slightly greater than that of the medium
\cite{altman2021holographic}.  An ellipsoid aligned with the optical
axis, $\theta = \pi/2$, scatters light in much the same way as a small
dense sphere.  The dependence of effective properties on ellipsoid
orientation is captured by the phenomenological relationship
\cite{altman2021holographic}
\begin{subequations}
  \label{eq:parametric}
  \begin{align}
    d_p^\ast(\theta) & = (d_{min} - d_{max}) \, \sin\theta + d_{max} \label{eq:thetadep} \\
    n_p^\ast(\theta) & = n_0 \left[ 1 + \frac{L}{d_p^\ast(\theta) - d_0} \right], \label{eq:charcurve}
  \end{align}
\end{subequations}
where $d_{max} = \SI{3.27}{\um}$ and $d_{min} = \SI{1.57}{\um}$ are
extracted from the maximum and minimum observed values of $d_p^*$.
Figure~\ref{fig:angledistribution}(a) presents experimental results
for \num{1712} colloidal ellipsoids obtained from the sample in
Fig.~\ref{fig:sem}.  Each data point reflects the effective diameter
and refractive index of a single ellipsoid captured at a random point
in its orientational trajectory.  The (red) dashed curve is a fit of
those data points to Eq.~\eqref{eq:charcurve} for
$L = \SI{65(2)}{\nm}$, $d_0 = \SI{1.24(1)}{\um}$ and
$n_0 = \num{1.40(0)}$.  The points then are colored according to the
inclination angle, $\theta$, of the closest point along that curve.

The random sampling of inclination angles in
Fig.~\ref{fig:angledistribution}(a) is compiled into a probability
distribution $P(\theta)$, that is plotted
Fig.~\ref{fig:angledistribution}(b).  Whereas the theory summarized in
Eqs.~\eqref{eq:jeffery} through \eqref{eq:jefferytheta} predicts that
Brownian ellipsoids are most likely to be aligned with the imaging
plane, $\theta = 0$, the measured distribution is clearly peaked
around $\theta = \pi/8$.  The same anomalous inclination has been
observed in measurements on colloidal dimers
\cite{altman2021holographic} and colloidal rods
\cite{zottl2019dynamics}, both of which are reproduced in
Fig.~\ref{fig:angledistribution}(b).

\begin{figure}
  \centering \includegraphics[width=\columnwidth]{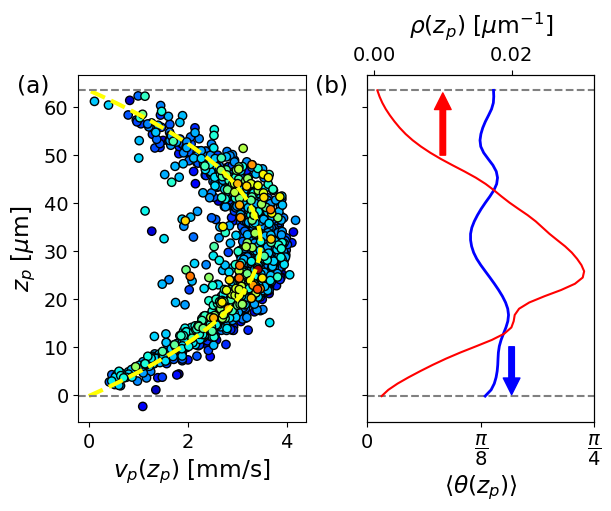}
  \caption{(a) Holographic tracking data obtained simultaneously with
    characterization results from Fig.~\ref{fig:angledistribution}(a).
    Each point indicates the axial position, $z_p$, and the flow
    speed, $v_p$, of a single particle and is colored by the
    inclination angle from Fig.~\ref{fig:angledistribution}(a).  The
    dashed (yellow) curve is a fit to the parabolic Poiseuille flow
    profile with a maximum speed of $v_0 = \SI{3.5}{\mm\per\second}$
    at the midplane.  Dashed (gray) lines represent estimates for the
    axial positions of the channel walls, $H = \SI{63.8}{\um}$
    obtained from the fit.  (b) The population-average inclination
    angle, $\avg{\theta(z_p)}$, does not depend significantly on the
    ellipsoids' height in the channel (blue curve) even though the
    probability distribution for particle positions, $\rho(z_p)$ shows
    a clear tendency for ellipsoids to travel near the channel's
    midplane.}
  \label{fig:z_theta}
\end{figure}

The qualitative discrepancy between the predicted and observed
distribution of inclination angles was noted in
\cite{zottl2019dynamics} and was emphasized in
\cite{altman2021holographic}.  This discrepancy is unlikely to result
from an experimental artifact because the same result is obtained with
orthogonal measurement techniques
\cite{altman2021holographic,zottl2019dynamics}.  It similarly cannot
be ascribed to uncertainty in the distribution of orbital constants,
$p(C)$, because each orbit described by Eq.~\eqref{eq:jeffery}
individually contributes to a peak in $P(\theta)$ at $\theta = 0$.

Lacking a definitive explanation for the observed anomalous
inclination, we review factors that are not included in the standard
formulation of Jeffery orbits that might affect tumbling transport of
aspherical particles in Poiseuille flows.  Figure~\ref{fig:z_theta}(a)
reports the axial position, $z_p$, and in-plane flow speed, $v_p$, for
each particle from Fig.~\ref{fig:angledistribution}(a).  The (yellow)
dashed curve is a fit to the parabolic Poiseuille flow profile.  The
resulting estimates for the channel height, $H = \SI{63(1)}{\um}$, and
midplane flow speed, $v_0 = \SI{3.5(2)}{\mm\per\second}$ are
consistent with the instrument's specifications.  The channel's height
is sufficiently large compared with the ellipsoids' major axis that
gradients in the shear rate are unlikely to have influenced the
distribution of observed inclination angles \cite{stover1990motion}.

The ellipsoids' shear Reynolds number in dodecane is smaller than
$\mathrm{Re}_s = \num{3e-4}$.  Although small, this may still have
been large enough for weak inertial effects to have influenced the
particles' orbital dynamics \cite{einarsson2015effect}.

Hydrodynamic coupling to the walls of the channel also may have
influenced the ellipsoids' trajectories.  This could explain the
nonuniform distribution of axial positions, $\rho(z_p)$, that is
plotted in Fig.~\ref{fig:z_theta}(b).  Redistribution of particles
away from the channel's walls and toward the midplane may be a
manifestation of hydrodynamic lift
\cite{ouchene2015drag,bagge2021parabolic}.  Hydrodynamic coupling to
the walls might also have a complementary effect on the ellipsoids'
orientational trajectories.  Any coupling-induced orientational bias
appears not to depend on position within the channel, however, because
the mean inclination angle, plotted in Fig.~\ref{fig:z_theta}(b), has
no obvious dependence on $z_p$.  This can be seen also in
Fig.~\ref{fig:z_theta}(a) because the distribution of particle
inclinations, denoted by the color of the plot symbols, shows no trend
with height.

\section{Discussion}

The trajectories of axisymmetric colloidal particles in simple shear
flows continue to present conundrums despite more than a century of
study.  Holographic particle characterization offers a fast and
effective way to amass large statistical samples that hopefully will
be useful for resolving some of these outstanding mysteries.
Analyzing holographic particle-characterization data in the
effective-sphere approximation yields useful estimates for elliptical
particles' out-of-plane orientations.  The same measurement also
yields each particle's position in the three-dimensional flow over a
comparatively large axial range, as well as the drift speed at that
position.

Holographic tracking of tumbling ellipsoids confirms previous reports
aspherical colloids do not behave as expected in plane Poiseuille
flows.  Rather than spending most of their time in the plane defined
by the flow and vorticity directions, these particles actually tend to
be inclined away from that plane.  The same angle of inclination,
$\theta = \pi/8$, is adopted by dimers with an aspect ratio of \num{2}
\cite{altman2021holographic}, rods with an aspect ratio as large as
\num{10} \cite{zottl2019dynamics}, and ellipsoids with an aspect ratio
of \num{5}.  The angle of inclination appears not to depend on the
particles' distance from bounding walls, even when the particles
themselves experience significant hydrodynamic lift.  Interparticle
collisions similarly are not likely to account for these anomalous
observations because the typical inter-particle separation exceeds the
channel height in all available studies.

\section*{Acknowledgements}

The work at New York University was supported by the National Science
Foundation through Award No.~DMR-2104837.  The xSight holographic
particle characterization instrument was acquired as shared
instrumentation with support fromthe MRSEC program of the NSF under
Award No.~DMR-1420073.  The Merlin FESEM as acquired through the
support of the MRI program of the NSF under Award No.~DMR-0923251.

We are grateful to Alexander Y.\ Grosberg and Aleksander Donev for
helpful conversations regarding Jeffery orbits. We also appreciate
useful comments from Andreas Zumbusch regarding the ellipsoidal
particle synthesis.

\section*{Competing Interests}

DGG is a founder of Spheryx, Inc., the company that manufactures the
instrumentation for holographic particle characterization that was
used for thus study.


%

\end{document}